\documentclass[superscriptaddress,secnumarabic,
amssymb,amsmath,nobibnotes,aps,prd,showkeys,showpacs,nofootinbib]{revtex4}%
\usepackage{graphicx}
\usepackage{epsf}
\usepackage{bm}
\usepackage{amsmath}
\usepackage{amsfonts}
\usepackage{amssymb}
\usepackage{epstopdf}
\usepackage{natbib}%
\providecommand{\U}[1]{\protect\rule{.1in}{.1in}}
\newcommand{\be}{\begin{equation}}
\newcommand{\ee}{\end{equation}}
\usepackage{color}

\begin{document}

\title{Spherically symmetric wormhole solutions in a general anisotropic matter field }

\author{Shibaji Halder}
\email{shibajihalderrrrm@gmail.com}
\affiliation{Department of Mathematics, Vidyasagar College, Kolkata-700006, India}
\author{Subhra Bhattacharya}
\email{subhra.maths@presiuniv.ac.in}
\affiliation{Department of Mathematics, Presidency University, Kolkata-700073, India}
\author{Subenoy Chakraborty}
\email{schakraborty@math.jdvu.ac.in}
\affiliation{Department of Mathematics, Jadavpur University, Kolkata-700032, India}
\begin{abstract}

The present work is an attempt to find possible traversable wormhole solutions in static spherically symmetric space-time supported by anisotropic matter field. Part of the work could be considered as a generalization of the work in Phys. Lett. B {\bf 757} (2016), 130 in the sense that it extends the work done therein. The paper provides several examples of wormholes in anisotropic pressure and provides general mechanisms for finding them. Finally the work examines the energy conditions corresponding to suggested wormhole solutions. 

\end{abstract}

\maketitle
\section{Introduction}
\label{Intro}

The pioneering work of Morris and Thorne \cite{1} has initiated the study of traversable wormholes (TWH) for spherically symmetric space time in the framework of Einstein gravity. For wormholes (WH) to be traversable it is necessary to have a red shift function without horizons or it is desirable to have a given asymptotic form for both the red shift and the shape function. Usually in the theoretical construction for WH geometry one assumes a priori the desired form of the red shift function and the shape function and the corresponding matter field is determined by the Einstein field equations. On the other hand, one may follow the conventional way of solving the Einstein field equations with prescribed matter field configuration and the red shift functions are obtained as solutions. In the first approach the energy density and pressure components are obtained as algebraic expressions which are mostly not realistic. For the alternative method one chooses barotropic equation of state $p_{r}=w_{r}\rho$ in order to solve the non linear field equations and often the traversablility of the WH or the consistency of the field equations demands that the matter be phantom in nature \cite{phmat}. 

The WH solutions in spherically symmetric space-time are mostly due to matter with anisotropic stress components. The examples of such anisotropic matter forming WH are in higher dimensional electromagnetic fields \cite{whelectro}, quadratic scalar fields \cite{kim}, anisotropic vacuum stress-energy of quantized fields \cite{quantscal}, scalar field minimally coupled to an electric field \cite{kimlee},  dark matter coupled to a quintessence scalar field (dark energy) \cite{dmde}, in non-commutative geometry modification \cite{kuh} and so on.

In recent past authors in ref. \cite{cat} have studied WH solutions for matter with isotropic pressure and have concluded few features on the corresponding space-time geometry. In this context the present paper studies WH solutions both with isotropic and anisotropic matter fields and infers on the structure of the space-time geometry for the formation of WH. It also provides general methods for obtaining WH solutions in both isotropic and anisotropic matter fields. In a sense it generalises and extends the results obtained in ref. \cite{cat}.

The paper is organised as follows: Section 2 deal with a general discussion on the most general class of TWH that was obtained by Morris and Thorne \cite{1} and shall serve as the WH objective for the current work. In section 3 we shall give general WH solutions for anisotropic matter. Solutions address anisotropic matter fields such that the ratio of radial to lateral pressure is a constant or a function of the radius. Section 4 is a discussion on a general method that can be used to obtain any WH solution in isotropic pressure corresponding to some valid WH shape functions. Finally the paper ends with a brief conclusion in section 5.

\section{Morris-Thorne traversable wormholes: A brief description }

Morris-Thorne \cite{1} described a spherically symmetric metric given by 
\begin{equation}
ds^{2}=e^{2\phi(r)}dt^{2}-\frac{dr^{2}}{1-\frac{b(r)}{r}}-r^{2}(d\theta^{2}+\sin^{2}\theta d\Phi^{2})\label{metric}
\end{equation}
as a possible solution to obtain viable WH. They imposed some restrictions on the unknown functions $\phi(r)$ and $b(r)$ so that the above metric describes a TWH successfully. It was hypothesised that in order for (\ref{metric}) to be a WH the function $b(r)$ should be such that it connects two asymptotically flat regions of space time with $b(r)$ attaining the minimum value at the junction called the throat. Since $b(r)$ was indicative of the shape at the WH throat, it was called the shape function.
Mathematically at the throat $r=r_{0}$ the shape function should satisfy:
\begin{equation}
b(r_{0})=r_{0},~
b'(r_{0})<1.\label{trt}
\end{equation} together with $b(r)<r$ for $r>r_{0}$ and $\frac{b(r)}{r}\rightarrow 0$ as $r\rightarrow\pm\infty$ which gives asymptotic flatness. $\phi(r)$ called the red-shift function is representative of the horizons, if any of the WH. It was found that WH with horizon would not be suitable for travel, hence it was concluded that the red-shift function should be finite at $r=r_{0}$ i.e. $(r-b(r))\phi'\rightarrow 0$ at the throat. Precisely $\phi=0$ results into a simple class of TWH solutions called the zero-tidal force WH. For a complete review see \cite{vis}. 

Assuming some known forms of the shape function and the red-shift function or in other words given any specific WH geometry, one could now determine the matter stress-energy tensor through the field equations, as given below:
\begin{align}
\kappa\rho(r)=&\frac{b'}{r^{2}}\\
\kappa p_{r}(r)=&-\frac{b}{r^{3}}+2\left(1-\frac{b}{r}\right)\frac{\phi'}{r}\\
\kappa p_{t}(r)=&\left(1-\frac{b}{r}\right)\left[\phi''+(\phi')^{2}+\frac{\phi'}{r}-\frac{b'r-b}{2r(r-b)}\phi'-\frac{b'r-b}{2r^{2}(r-b)}\right]
\end{align}
This also meant that the constraints on the shape function $b(r)$ and the red-shift function $\phi(r)$ that was used to fix a tenable geometry to the WH that would be suitable for human travel, in turn now imposed restrictions on the matter stress-energy tensor. Mathematically it was found that the constraint of minimum radius at the throat $r_{0}$ together with the traversability criterion imposed enormous tension at the throat. To a static traveller this would appear as negative energy density with the resulting matter violating null energy conditions (NEC) \cite{vis, nec} at the throat. That is to say:
\begin{equation}
\kappa(\rho+p_{r})=\frac{b'r-b}{r^{3}}+2\left(1-\frac{b}{r}\right)\frac{\phi'}{r}<0
\end{equation}
This was an essential feature of the Morris-Thorne TWH. It is evident that in order to successfully construct a TWH model, one needs to fix five unknowns namely, $b(r),~\phi(r),~\rho(r),~p_{r}$ and $p_{t}.$ For that one had the three field equations and one energy conservation equation $T^{\mu\nu}_{;\nu}=0$ given by
\begin{equation}
p'_{r}=\frac{2}{r}(p_{t}-p_{r})-(\rho+p_{r})\phi'(r).\label{consv}
\end{equation} 
If matter is considered to be isotropic, i.e. 
\begin{equation}
p_{r}=p_{t}
\end{equation} then (\ref{consv}) reduces to $p_{r}'=-(\rho+p_{r})\phi'(r).$ For zero-tidal force TWH, i.e. for $\phi(r)=0$ one would obtain $b(r)=\frac{r^{3}}{r_{0}^{2}}$ a shape function that is not suitable for describing the geometry of an asymptotically flat TWH with throat situated at $r_{0}.$ It is thus evident that isotropic matter cannot in general support zero tidal force WH \cite{cat}.

\section{Morris and Thorne wormholes with anisotropic matter}  

Here we consider the WH solutions in anisotropic matter stress energy tensor, i.e. we consider
\begin{equation}
p_{t}=\alpha p_{r}\label{prel}.
\end{equation}
Where $\alpha,$ the anisotropic parameter can be a constant ($\neq 1$) or it can be a function of the radius $r.$ Then using (\ref{prel}) in the last two relations of the field equations we arrive at the following master equation 
\begin{equation}
rb'(r)(r\phi'+1)+b(r)[2r^{2}(\phi''+\phi'^{2})-r\phi'(4\alpha-1)-(2\alpha+1)]-[2r^{3}(\phi''+(\phi')^{2})+2r^{2}\phi'(1-2\alpha)]=0.\label{De}
\end{equation}
The above equation has three unknowns $b(r),~\phi(r)$ and $\alpha.$ Hence using (\ref{De}) it is possible to freely determine one unknown  based on certain restrictions on the other two functions.

\subsection{Constant anisotropy parameter, $\alpha=constant (\neq 1)$}

{\bf Method I}
\vspace{2em}

Corresponding to constant anisotropy parameter our first approach would be to solve for unknown $b(r).$ Interpreting (\ref{De}) as a first order differential equation in unknown $b(r)$, we employ the following technique to obtain solutions.
\subsubsection*{Case 1}

We assume 
\begin{equation}
r(\phi''+\phi'^{2})+\phi'(1-2\alpha)=0.\label{c1}
\end{equation}
Solving the above for $\phi(r)$ gives
\begin{equation}
e^{\phi(r)}=\phi_{1}+\frac{\phi_{0}}{2\alpha}r^{2\alpha}\label{phi1}
\end{equation}
with $\phi_{0},~\phi_{1}(\neq 0)$ being two arbitrary constants. Using the above expression for $\phi(r)$ we now solve (\ref{De}) and obtain $b(r)$ as follows:
\begin{equation}
b(r)=\frac{b_{0}^{2\alpha}r^{2\alpha+1}}{\left[\frac{\phi_{0}}{2\alpha\phi_{1}}(2\alpha+1)r^{2\alpha}+1\right]^{\frac{2\alpha}{2\alpha+1}}},\label{bt1}
\end{equation}
$b_{0}$ the constant of integration. This will be admissible as WH shape function provided it satisfies the throat condition (\ref{trt}), i.e. (\ref{bt1}) is now given by:
\begin{equation}
\frac{b(r)}{r}=\left(\frac{r}{r_{0}}\right)^{2\alpha}\left(\frac{\frac{\phi_{0}}{2\alpha\phi_{1}}(2\alpha+1)r_{0}^{2\alpha}+1}{\frac{\phi_{0}}{2\alpha\phi_{1}}(2\alpha+1)r^{2\alpha}+1}\right)^{\frac{2\alpha}{2\alpha+1}}.\label{b1}
\end{equation}
Further asymptotic flatness requirement demands $\alpha<0.$ Hence for $\alpha<0$ solutions (\ref{phi1}) and (\ref{b1}) represented spherically symmetric asymptotically flat space-time geometries that could serve as a WH. Evaluating the stress-energy momentum tensor at the throat showed
\begin{equation}
\kappa\rho=\frac{b}{r^{3}}\left(1+2\alpha\frac{re^{\phi}}{\phi_{0}r^{2\alpha}+e^{\phi}}\right)>0\label{rho1}
\end{equation}
for $-\frac{1}{2c}<\alpha<0$ where $c=\frac{r_{0}e^{\phi(r_{0})}}{\phi_{0}r_{0}^{2\alpha}+e^{\phi(r_{0})}}$ is a positive number. Accordingly the radial pressure $p_{r}(r_{0})=-\frac{1}{r_{0}^{2}}<0$ and $p_{t}(r_{0})>0.$ Evidently the NEC at the throat given by $\kappa(\rho+p_{r})=\frac{2c\alpha}{r_{0}^{2}}$ is violated therein for the given range of the anisotropy parameter. Further glancing at the equation of state parameter $\omega_{r}$ corresponding to the radial pressure we find that, for the given range of $\alpha,~\omega_{r}=\frac{p_{r}}{\rho}<-1$ at the throat $r_{0}.$ This means that class of WHs described above is supported by dark energy fluid or super-quintessence/phantom fluid. Also, one can obtain the zero tidal force WH corresponding to the choice $\phi_{0}=0$ and $\phi_{1}=1,$ which are indeed the Morris-Thorne TWH.

\subsubsection*{Case 2}

Here we assume
\begin{equation}
2r(\phi''+\phi'^{2})-\phi'(4\alpha-1)=0
\end{equation}
which gives 
\begin{equation}
e^{\phi}=\phi_{1}+\frac{\phi_{0}}{2\alpha+1/2}r^{2\alpha+1/2}.
\end{equation} 
Corresponding to the above results we obtain $b(r)$ as follows 
\begin{equation}
b(r)=b_{0}\frac{2\gamma r^{\gamma+1}}{\left[\frac{\phi_{1}\gamma}{\phi_{0}}+(\gamma+1)r^{\gamma}\right]^{\frac{\gamma+1/2}{\gamma+1}}}\left(\frac{\phi_{1}\gamma}{\phi_{0}}\right)^{-1/(2\gamma+2)}   ~_2F_{1}\left(1/2\gamma,1/(2\gamma+2),1+1/2\gamma,-\frac{\gamma+1}{\frac{\phi_{1}\gamma}{\phi_{0}}}r^{\gamma}\right)
\end{equation}
where $\gamma=\alpha+1/4$ and $~_2F_{1}$ is the Hypergeometric function for $|r|<1.$ Existence of asymptotically flat spherically symmetric WH solutions cannot be established using the above complicated form of $b(r)$ and is not useful for practical applications.

\vspace{2em}
{\bf Method II}
\vspace{2em}

Our second approach is to solve for (\ref{De}) corresponding to unknown $\phi(r).$ Accordingly we rewrite (\ref{De}) as:
\begin{equation}
\phi''+(\phi')^{2}+\frac{b'r-(4\alpha-1)b-2r(1-2\alpha)}{2r(b-r)}\phi'+\frac{b'r-(1+2\alpha)b}{2r^{2}(b-r)}=0\label{DEphi}
\end{equation}
In order to solve (\ref{DEphi}) we make the simplifying assumption that $b'r-(1+2\alpha)b=0.$ This gives $b(r)=b_{0}r^{2\alpha+1}.$ Again  applying the throat constraints we get
\begin{equation}
\frac{b(r)}{r}=\left(\frac{r}{r_{0}}\right)^{2\alpha}\label{b2}
\end{equation}
with $\alpha<0.$ Using (\ref{b2}) we solve (\ref{DEphi}) and obtain $\phi(r)$ as given:
\begin{equation}
e^{\phi}=\phi_{1}+\phi_{0}\sqrt{1-\left(\frac{r}{r_{0}}\right)^{2\alpha}}\label{phi2}
\end{equation}
with $\phi_{0},~\phi_{1}(\neq 0)$ being arbitrary constants. (\ref{b2}) and (\ref{phi2}) together provide another example of spherically symmetric WH geometry that is possible in anisotropic matter. Considering the matter stress-energy tensor at the throat we obtain
\begin{equation}
\kappa\rho=\frac{b}{r^{2}}(2\alpha+1)>0\label{rho2}
\end{equation}
for $-\frac{1}{2}<\alpha<0$ and the radial pressure at the throat is accordingly given by $\kappa p_{r}=-\frac{1}{r_{0}^{2}}<0.$ Again the NEC are not satisfied at the throat $r_{0}.$ Further the equation of state parameter corresponding to the radial pressure is such that $\omega_{r}<-1$ which clearly show that matter supporting the throat will be dark energy or phantom in nature as obtained before. 

It is interesting to note that corresponding to $\phi_{0}=0,$ WH solutions given by (\ref{phi1}), (\ref{b1}) and (\ref{b2}), (\ref{phi2}) are same. This shows that the only class of zero-tidal force TWH that are possible in anisotropic pressure with constant pressure function is essentially given by the shape function $\frac{b(r)}{r}=\left(\frac{r}{r_{0}}\right)^{2\alpha}$ with $-\frac{1}{2}<\alpha<0$ and the matter make up of such TWH being phantom in nature.

\subsection{Variable anisotropy parameter, $\alpha=\alpha(r)$}

Here we consider $\alpha$ to be a function of the radial coordinate $r.$ For variable $\alpha$ the differential equation (\ref{De}) has three unknown functions. In order to make the equation tractable, we shall consider only the class of zero-tidal force WH, i.e. $\phi(r)=constant.$ Then from (\ref{De}) one obtains
\begin{equation}
\frac{b(r)}{r}=b_{0}e^{\int\frac{2\alpha(r)}{r}dr}\label{bt4}
\end{equation}
with $b_{0}$ being the constant of integration. In order to determine valid WH geometry we need to assign specific form to the anisotropic parameter $\alpha.$ Accordingly we take the simplest possible form of $\alpha=\alpha_{0}r^{\beta},$ where $\alpha_{0}$ and $\beta$ are some constants. A corresponding viable shape function is obtained for $\frac{2\alpha_{0}}{\beta}<0$ and is given by:
\begin{equation}
\frac{b(r)}{r}=e^{\frac{2\alpha_{0}}{\beta}(r^{\beta}-r_{0}^{\beta})}.\label{b4}
\end{equation}
Again the corresponding matter stress-energy tensor at the throat $r_{0}$ gives
\begin{equation}
\kappa\rho=\frac{b(r)}{r^{3}}(2\alpha_{0}r^{\beta}+1)>0
\end{equation}
for $-\frac{1}{2r_{0}^{\beta}}<\alpha_{0}<0$ and $\beta>0.$ Accordingly for the above given values of $\alpha_{0}$ and $\beta$ the NEC is violated at the throat. Analysing the equation of state parameter $\omega_{r}$ we find that $\omega_{r}<-1$ for the given range of $\alpha_{0}.$ This shows that the class of zero-tidal force solutions obtained is supported by phantom energy. Thus another class of zero-tidal force TWH solutions for anisotropic pressure can be obtained, which are altogether new and completely different from the previously obtained power law shape function, that is common in the literature.

\section{Solution corresponding to isotropic pressure}

In this section we shall consider the equation (\ref{De}) corresponding to $\alpha=1.$ That is we are considering isotropic pressure solutions to (\ref{De}):
\begin{equation}
\phi''+\phi'^{2}-\frac{b'r-3b+2r}{2r(r-b)}\phi'-\frac{b'r-3b}{2r^{2}(r-b)}=0.\label{isode}
\end{equation}
To solve the above differential equation we make the substitution $\phi'(r)=\psi(r),$ which reduces it to
\begin{equation}
\psi'(r)=\frac{b'r-3b}{2r^{2}(r-b)}+\frac{b'r-3b+2r}{2r(r-b)}\psi-\psi^{2},\label{ric2}
\end{equation}
the generalized Riccati equation in unknown $\psi(r).$ We transform this into a second order homogeneous equation in unknown $u(r)$ by taking the substitution $\psi(r)=\frac{u'(r)}{u(r)}$ as given below:
\begin{equation}
\frac{d^{2}u}{dr^{2}}-\frac{b'r-3b+2r}{2r(r-b)}\frac{du}{dr}-\frac{b'r-3b}{2r^{2}(r-b)}u=0.\label{serde}
\end{equation}
This can be solved for any known $b(r)$ that full fills the WH geometry requirements. We make the simple choice of $\frac{b(r)}{r}=\frac{1}{1-r_{0}+r}$ that satisfies the constraints imposed on the WH throat $r=r_{0}$. (Note that we are avoiding the power law form because, firstly known solutions corresponding to such shape functions already exist in the literature and secondly existence of some unknown $\beta$ as exponent of $r$ can complicate the solution and make it less tractable). Using the above $b(r)$ we can obtain a solution to $\phi(r)$ as given below:
\begin{equation}
e^{\phi}=\sum_{n=0}^{\infty}[\phi_{0n}\sqrt{r-r_{0}}+\phi_{1n}](r-r_{0})^{n}, r_{0}<r<r_{0}+R\label{phiser}
\end{equation}
for some radius $R,$ and
\begin{align*}
\phi_{01}&=-\frac{4-r_{0}^{2}}{6r_{0}^{2}}\phi_{00},\\
\phi_{02}&=-\frac{(2+r_{0})(2+3r_{0})\phi_{01}+4(1-r_{0})\phi_{00}}{20r_{0}^{2}},\\
\phi_{0n+3}&=-\frac{(2n+1)(2n-3)\phi_{0n}+4[n(2n+3)(2r_{0}+1)+3]\phi_{0n+1}+(4+2r_{0})[r_{0}(2n+3)(2n+5)+2]\phi_{0n+2}}{4r_{0}^{2}(2n+7)(n+2)},n\geq 0\\
\phi_{11}&=-\frac{2+r_{0}}{r_{0}^{2}}\phi_{10},\\
\phi_{12}&=-\frac{(2+r_{0})\phi_{11}+3\phi_{10}}{6r_{0}^{2}},\\
\phi_{1n+3}&=-\frac{n(n-2)\phi_{1n}+[n(2n-1)(2r_{0}+1)+3]\phi_{1n+1}+(2+r_{0})[2r_{0}(n+1)(n+2)+1]\phi_{1n+2}}{r_{0}^{2}(n+3)(2n+3)},n\geq 0
\end{align*} 
with $\phi_{00}$ and $\phi_{10}$ being arbitrary. Further it is found that at the throat the matter stress-energy tensor satisfies $\kappa\rho>0$ for $0<r_{0}<1,$ and for this value of $r_{0}$ it is seen that the NEC is violated at the throat and that $\omega_{r}<-1$. Thus the above solution provides the most general class of spherically symmetric WH solution supported by isotropic fluid that is essentially phantom in nature, corresponding to the above chosen shape function. 

In fact we stress that the above class of solution is one of the most general WH solution that can be obtained corresponding to a perfect fluid stress-energy tensors. The above solution also provides a general method for obtaining a wide class of solutions corresponding to any known shape function. Further one can apply the above method to obtain successful WH solution corresponding to anisotropic matter with any known value of constant anisotropic parameter.

\section{Conclusion}

In this paper we have discussed the existence of WH in anisotropic pressure conditions. We have discussed general methods for obtaining spherically symmetric WH solutions that are Morris-Thorne type. Further, spherically symmetric WH solution supported by a phantom like isotropic fluid with a general shape function (i.e. not the power law shape function) is provided in details. It could be further shown that the WH solution provided in the paper corresponding to anisotropic fluid can become traversable for specific choice of arbitrary integration constants. Since TWH has been speculated as means of time travel \cite{tt}, solutions describing such TWH geometry has been in prominence with the advent of Morris-Thorne type WH. Our solutions provide two such general class of TWH geometry in anisotropic pressure that is supported by the phantom energy fluid or by dark energy fluid.  Several works on such phantom energy WH and their importance could be found in \cite{phwh}.

Through our calculations we have also shown that zero-tidal force TWH can exist only in anisotropic fluid, with supporting matter essentially dark energy type or phantom in nature. The above work asserts the conclusion of \cite{cat} that zero tidal force WH are not possible in isotropic fluid. Further, it also suggests a general method of obtaining WH solutions in isotropic pressure corresponding to any given WH shape function. So far the literature on WHs concentrate on examining specific WHs and their corresponding energy condition violations at the throat. The current work on the other hand generalises this entire process of finding wormholes under any matter field. The suggested solutions are the most general ones and any WH existing in the literature can be obtained from them by proper choice of parameters.

\section{Acknowledgments}
 SB acknowledges UGC's Faculty Recharge Programme and Department of Science and Technology, SERB, India for financial help through ECR project (File No. ECR/2017/000569). SC thanks IUCAA, Pune, India, for their warm hospitality while working on this project.


\end{document}